\documentclass[a4paper]{article}

\usepackage{amssymb,amsfonts}       

\usepackage{epsfig}
\usepackage{amsmath}
\usepackage{rotating}
\begin{document}

\title{\bf A cyclic universe with colour fields}

\author{V. N. Yershov \\
{\small \it Mullard Space Science Laboratory (University College London),} \\ 
{ \small \it  Holmbury St.Mary, Dorking RH5 6NT, UK} \\
{ \small vny@mssl.ucl.ac.uk}}

\date{}

\maketitle

\begin{abstract}
The topology of the universe is discussed in relation to the 
singularity problem. We explore the possibility that the initial 
state of the universe might have had a structure with 3-Klein bottle
topology, which would lead to a model of a nonsingular oscillating 
(cyclic) universe with a well-defined boundary condition. 
The same topology is assumed to be intrinsic to the nature 
of the hypothetical primitive constituents of matter (usually called preons) 
giving rise to the observed variety of elementary particles. 
Some phenomenological implications of this 
approach are also discussed. 
\end{abstract}

\newcommand{\rum}{\rule{0.5pt}{0pt}}
\newcommand{\rub}{\rule{1pt}{0pt}}
\newcommand{\rim}{\rule{0.3pt}{0pt}}
\newcommand{\numtimes}{\mbox{\raisebox{1.5pt}{${\scriptscriptstyle \times}$}}}

\renewcommand{\refname}{References}
\renewcommand{\figurename}{\small Fig.}

\newcommand{\abs}[1]{\mid\negmedspace#1\negmedspace\mid}

\section{Introduction}
\label{introduction}

During the past few years the old idea of a cyclic universe
\cite{lemaitre33}
has regained new ground following a series of publications by 
Stein\-hardt and Turok 
\cite{steinhardt02a} 
describing a model based on 10-dimensional colliding branes. This model 
sur\-mo\-unts the hurdles of the previous cyclic-universe models, such as 
the problem of growing entropy \cite{markov84} and the lack of a realistic physical 
mechanism for the rebounding of the oscillating universe after each 
collapse \cite{guth83}.
That is why the Steinhardt-Turok model has attracted much attention \cite{bojowald04}.
However, it requires the universe to have at least ten spatial dimensions,
which is not yet confirmed by observational evidence. 
Here we shall return to a three-dimensional cyclic-universe model,
in which the distinction between matter and space is abandoned,
as was proposed by Wheeler \cite{wheeler62} and some other authors \cite{ambjorn06}.
That is, we shall regard the matter particles as stable configuration 
patterns of a moving manifold (space), assuming that the notion of geometry is 
dynamical and that the properties of matter are underlied by the geometry of 
spacetime, which is the approach accepted by many 
authors \cite{rovelli04}.

The main purpose of this paper is to show that the origin of the observed variety 
of particle species might be related to the global properties of 
spacetime and that the dynamics of the universe as a whole might be
determined by the properties of the basic building blocks of matter.
We shall focus mainly on the topological aspects of the 
problem, gaining insight into the nature of the cosmological 
singularity by hypothesising that the initial state of the 
universe might have had a structure with 3-Klein bottle topology.
We also expect this approach to shed light on the singularity problem 
related to microscopic and macroscopic objects, such as elementary particles
and collapsed stars.

Astronomical observations provide strong evidence that compact objects with 
sufficiently large masses for contraction into black holes  
do exist in nature \cite{narayan05}, e.g., in compact binary systems with 
calculable masses of the components where a few dozens of low-mass (4 to 20 solar masses)
black hole candidates were found \cite{snezhko72}, in ultra-luminous compact X-ray sources 
\cite{charles01} whose luminosities exceed the Eddington luminosity of a 
100-solar-mass object, and in galactic nuclei \cite{eckart96} 
where the physical processes and the dynamics of stars near the nuclei indicate 
that the masses of the central objects exceed $10^6$ solar masses \cite{beifiori07}
All this suggests thus that black holes may exist as 
real physical objects (see the reviews \cite{novikov89} and references therein).
The singularities theorems \cite{penrose69} require black holes 
to have singularities. 
There exists also observational evidence \cite{gamow48} suggesting that
our universe in the past was contracted to a very small volume,
which is regarded by the standard cosmological model as an indication 
that the universe could be born from a singularity.  
On the other hand, singularities could, in principle, be avoided 
by relaxing the conditions required by the singularities theorems
\cite{dymnikova02}. 
For example, within the classical 
framework of Einstein--Cartan theory \cite{cartan22}, torsion 
contributing to the energy--momentum tensor leads to an effective 
negative pressure and eliminates the singularity.
Yet, most physicists believe that general relativity is incomplete 
because it ignores quantum effects and that it has to be replaced 
with another theory (quantum gravity) to deal with singularities 
\cite{ashtekar05}. Then the singularity must be ``smeared'' by quantum 
fluctuations \cite{narlikar79}. But still there are some hints that 
this might not be the case: specially organised astronomical observations 
\cite{ragazzoni03} detect no quantum fluctuations in vacuum which are 
needed to smear the singularity. 
It has also been indicated in the literature \cite{mitrofanov03} that
the effects of quantum gravity, which have had $10^{10}$ years
to act on the gamma-ray burst photons, should completely randomise 
the polarisation of these photons, contrary to what is 
observed \cite{dado07}.  

%
Following these clues, we shall explore here the possibility
that general relativity can be valid 
up to a near the Planck-length scale, 
which would correspond to the ideas developed
by Markov \cite{markov82} and others \cite{aman84}
about the existence of an upper bound on spacetime curvature.
Implementing this idea, 
we shall replace a singularity with a topological feature 
of the manifold (see, e.g., the discussion of this issue
by Ellis \cite{ellis73}). We assume that such a replacement 
is possible because non-singular solutions to the Einstein field 
equations are known to exist. 
They include, in particular, a nonsingular topological feature
known as the Einstein-Rosen bridge \cite{einstein35},
as well as some solitonic solutions with scalar fields 
used for modelling particles \cite{bronnikov73}, the false-vacuum 
bubbles \cite{fischler90}, Chern-Simon vortices \cite{ibanez83} 
and Falaco solitons \cite{kiehn01}.
There are many models that consider topological features 
as particles or systems of interacting particles \cite{finkelstein59}. 
We shall assume here that such systems can, indeed, be related to 
the internal structures of elementary particles (the fundamental fermions). 
In other words, we shall treat these particles as composite entities.
The evidence of the compositeness of the fundamental fermions
is mostly indirect: it can be seen in the triplication of their 
species, classified by the standard model of particle physics 
in three generations with almost identical properties except for the 
hierarchical pattern of their masses and mixings \cite{ellis82}. 
Historically it is known that patterns in particle properties 
were always related to some underlying structures. 
There exists also some experimental (albeit inconclusive) evidence of 
quark compositeness,  which comes from proton-proton 
and positron-proton scattering experiments \cite{eichten83}.  
These experiments show that the probability of particle scattering for 
the most energetic collisions (above 200 GeV) 
is significantly higher than that predicted by current theoretical models. 
The experiments with quark-quark scattering by the Collider 
Detector at Fermilab (CDF group)  \cite{abe93} also 
showed evidence for substructure within the quark. 
Even though the later measurements made by the D0 collaboration 
\cite{abbott01} did not confirm the excess of the scattering 
probability for high  energy jets, the results of all these 
experiments, taken in the context of the observed 
pattern of quark properties, make a strong point in favour of quark 
compositeness. Perhaps the difficulties with detecting the scattering
probability excess could be attributed to the inaccessibility of the 
compositeness scale, which must correspond to $10^2$ to $10^3$\,\,TeV or 
larger, according to the estimations based on the observed anomalous magnetic 
moments of leptons  \cite{robinett83}.

Most of the existing composite models describe each quark and 
lepton as a combination of three simpler entities called preons \cite{dsouza92}.  
In the preon model of Salam and Pati \cite{pati73} a quark or 
lepton contains one of the three preon types called ``somons'' 
that determines its generation, one 
of two ``flavons'' that determines its flavour and electric 
charge, and one of four ``chromons'' that determines its colour and 
modifies its electric charge. Somons are electrically 
neutral and colourless. Flavons have electric charges of 
either +1/2 or -1/2 of the electron's charge and are colourless. 
Chromons that are red, green, or blue have a charge of +1/6, 
while the colourless chromon has a charge of -1/2. The 
possible combination of $3 \times 2 \times 4$ preons gives 
all the 48 quarks and leptons with their appropriate generations, 
colours, and charges, 
but the choice of the preon quantum numbers for this model
remained unexplained.

A more congruous version of the preon model is the rishon 
model of Harari and Seidberg \cite{harari79}, 
which describes all particles in a particular generation as 
three-particle combinations of ``rishons''. There are two rishon 
types, each type having three possible colours and hypercolours, 
with generations as excited states of the three-rishon system, 
so that the rishon model uses only 2 preons and their antimatter 
counterparts to generate the 48 quarks and leptons.

Yet another model with only two kinds of preons is the trion model
proposed by Raitio \cite{raitio80} describing the first generation of quarks 
and leptons as three-body combinations of preons with colour and sub-colour 
charges besides the usual electric charge. 
Raitio makes use of the idea that preons must be of the minimal 
possible size and maximal possible mass (calling them ``maxons''), 
which resemble in a sense  Markov's primitive particles called "maximons"
\cite{markov65}.
In a more recent preon model proposed by Bilson-Thompson, 
Markopoulou and Smolin \cite{bilson07}, which was likely 
inspired by the more general topological theory developed in 1999 
by Khovanov \cite{khovanov99}, 
the fundamental fermions appear as invariant states corresponding 
to the braidings of manifolds with genus 3, 
each of the possible braidings having been identified with a 
standard model particle. This model is quite different from the 
other preon models by representing preons as extended objects
with complicated topology arising from spacetime dynamics. 
But using multiple-genus manifolds for this purpose is actually 
equivalent to using multiple basic entities for the construction 
of the standard model particles because the genus number is not 
emerging naturally from the model but has to be chosen such as 
to generate the appropriate variety of the standard model particles. 
This idea has been used in a newer topological 
preon model by Mongan \cite{mongan08} who has combined 
Bilson-Thompson's wrappings with the holographic 
principle and represented the fundamental fermions as being
formed of three distinguishable preon strands bound by non-local 
three-body interactions.

Despite the evidence of the compositeness of the
fundamental fermions, the preon  models remain unpopular, 
mainly because these models face numerous problems,
one of which is the problem of the preon mass. From 
scattering experiments it is known that the hypothetical 
compositeness scale corresponds to the distances smaller than $10^{-18}$~m. 
The momentum uncertainty of a particle confined  
within the region of this size is about 200~GeV, which
is much larger than the masses of the first family quarks.
This difficulty can be overcome by postulating a new force, which would be
many orders of magnitude stronger than the known strong force. 
This would add a considerable complication to 
the standard model, but with such a hyperforce the preons would be tightly bound 
inside a quark, and the energy from their large momentum 
would be cancelled by their large mass defect (binding energy). 
This approach is quite promising, 
and here we shall adhere to it, disregarding the fact that so far none 
of the previous attempts to use it for the explanation of the quark and 
lepton properties has succeeded. 

Not only does the mass problem make the composite models of fermions 
unpopular, but also the fact that these models face grave problems 
with gauge anomalies and diverging energies on small scales
\cite{weinberg76}.
Gauge anomalies (undesirable symmetry breaking)
appear to be due to unavoidable approximations in (perturbative) models based 
on quantum field theory. Here we shall avoid this problem by using a 
general relativistic (classical) platform, for the classical fields are 
known to be intrinsically anomaly-free.

The main difference between the model to be presented here and the 
other preon-based models is that here we shall go all the way down in 
reducing the number of primitive particle types.
As we have already mentioned, all of the previous composite models 
explain the observed variety of 
elementary particles by different combinations of a certain number of preon 
types, reduced with respect to the number of the fundamental types 
in the standard model \cite{pati74}. 
However, it is plain to see that even this reduced number 
cannot solve the problem. It is not worthwhile replacing one 
variety of the basic entities with another, if the origin 
of these new varieties remains unexplained. Only a model based 
on a {\sl single} entity would make any sense, which is what 
we are going to propose here. Namely, we suggest that there should be a 
single preon type, having no flavours, spins or any other quantum numbers, 
and carrying only the electric and colour charges.
That is, the preons in our model will be represented by unit charges 
with the SU(3)$\times$U(1)-symmetry of their fields.

Our model will differ from the other preon-based models by 
yet another aspect: within its framework it will be possible to  
analyse preon dynamics, whereas the authors of previous models 
acknowledge the lack of dynamics as one of 
the major drawbacks in their models. 
Even in the case of the Bilson-Thompson's preons,
emerging from the quantum dynamics of spacetime, 
the authors admit that the effective low-energy preon 
dynamics is unknown because it is completely decoupled from the underlying 
spacetime dynamics \cite{bilson07}.  
In our view, it is the preon dynamics that is crucial for understanding
the properties of the standard model particles. By not going 
beyond simple combinatorial considerations one would never be able to uncover the 
origin of, for example, particle magnetic moments. And 
most of the previous preon models do not address this important 
property of the fundamental fermions.  

We shall outline the proposed approach in the next two sections. 
Then, in Sections\,\,4 and 5, 
we shall demonstrate the functionality of this model by presenting a few 
simple examples of preonic
bound states. In Section\,\,6 we shall comment on some possibilities
of experimental verification of our model, and in Section\,\,7 we shall discuss 
some implications of this approach in connection with the problems of cosmological 
singularity and oscillating universe models.

\section{Primitive particle}
\label{primitiveParticle}

\begin{figure}[htb]
\begin{turn}{-90}\epsfig{figure=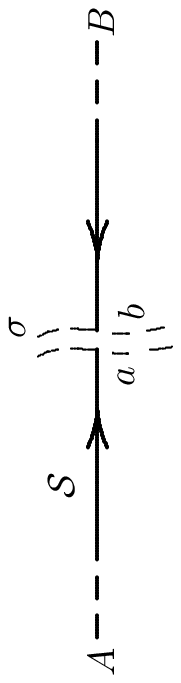,width=3.0cm}\end{turn}
\put(100,-0.1){
 \makebox(0,0)[t]{\begin{turn}{-90}\epsfig{figure=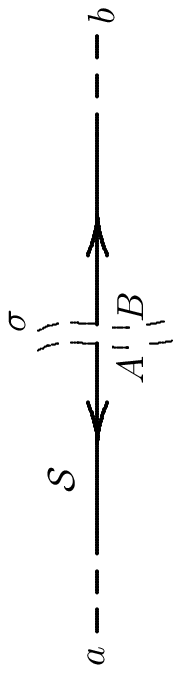,width=3.0cm}\end{turn}
}}
\caption{A point-like discontinuity $\sigma$ on a (locally flat) 3-manifold 
$\mathcal{S}$ with the boundaries of $\mathcal{S}$ labelled as $AB$ and $ab$.
The inflow (left) or outflow (right) of test particles is regarded as a charge
of $\sigma$.} 
\label{fig:blackhole}
\end{figure}

Consider a spinning 3-manifold (e.g. a sphere), $\mathcal{S} \in \{\mathbb{S}^3\}$, 
of radius $R_\mathcal{S} \in (-\infty, +\infty)$ formed of a massless 
elastic fluid -- a collection of worldlines of test particles. 
The rotation of $\mathcal{S}$ 
corresponds to a 4-velocity $\mathbf{u}$ of the fluid in all possible 
directions through each point of the spatial slice of the manifold 
(for a discussion of spinning 3-manifolds see, e.g., \cite{artin26}). 
Let the manifold be punctured, $\mathcal{S} \rightarrow \mathcal{S}\backslash\{0\}$, 
that is, containing a point-like discontinuity $\sigma$, which we shall use 
to represent a singularity (Fig.\ref{fig:blackhole}) and which we shall regard 
as a primitive particle that has no properties except for the property of 
possessing a charge (inflow or outflow of test particles, as indicated by 
arrows in Fig.\ref{fig:blackhole}).

\noindent
Given a compact control surface $\mathcal{C}$ on $\mathcal{S}$ 
($\mathcal{C} \in \mathbb{S}^2$) with no singularities
 in its interior, the inflow through this surface should be  
equal to the outflow (Fig.\ref{fig:conservation}), which is the usual 
flow  conservation condition. In the case of a singularity $\sigma$
enveloped by $\mathcal{C}$, either inflow or outflow vanishes, 
which is not consistent with the conservation principle.
To restore consistency, the flow should be sent back to $\mathcal{S}$,
which can be done through the boundaries by identifying $ab$ with $AB$:
\begin{figure}[htb]
\begin{turn}{-90}
\centering 
\epsfysize=2cm
\includegraphics[scale=0.8]{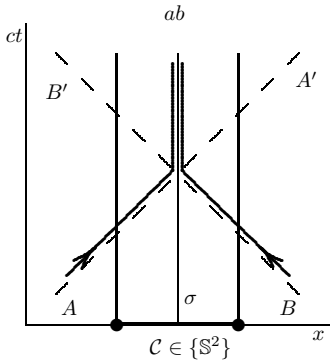}
\end{turn}
\caption{Worldlines of flow particles $AA'$, $BB'$ (no singularity) and 
$Aa$, $Bb$ (entering a singularity $\sigma$ wrapped with a control 
2-sphere $\mathcal{C}$). In the latter case the flow is not conserved.}
 \label{fig:conservation}
\end{figure}
\begin{equation}
\begin{matrix}
{\rm a~\rightarrow~A} \\
{\rm b~\rightarrow~B} 
\end{matrix}
\label{eq:boundaries1}
\end{equation}
or by crisscrossing them:
\begin{equation}
\begin{matrix}
{\rm a~~~A} \\
 \hspace{-1.4mm} {\vspace{-0.9mm} \hspace{0.4mm}\nearrow 
 \hspace{-3.9mm} {\vspace{1.2mm} \searrow}}  \\
{\rm b~~~B} 
\end{matrix}
\label{eq:boundaries2}
\end{equation}

\noindent
The resulting shape is a hypertorus $\mathbb{T}^3$ (for the first case), 
or the Klein bottle $\mathbb{K}^3$ (for the case with the crisscrossed
boundaries), which are known to be reasonable simple manifolds for 
cosmological models \cite{ellis71}. This restores the integrity of the flow and 
replaces the discontinuity at the point $\sigma$ by a topological feature
-- the central opening  (``throat'') of the hypertorus
or the Klein bottle. 
The inflow (outflow) $Q$ through this object 
is still not accounted for, but the total flow of the system 
is conserved. 
As we have already mentioned in Section\,1, such a topological
feature corresponds to the non-singular particle-like solution 
of Einstein's equations describing the Einstein-Rosen bridge 
(wormhole) whose properties have been studied by many 
authors \cite{morris88}. It is worthwhile mentioning the analytic 
solution obtained recently by Shatskii, Kardashev and Novikov \cite{shatskii08},
which describes an infinite number of wormhole-connected spherical 
universes. This structure is interpreted by its authors as a multitude 
of simultaneously existing independent worlds (a multiverse) emerging 
from quantum vacuum in 
different regions of the spacetime manifold. From the viewpoint 
of the colour-preon model this solution corresponds to a single 
universe containing an infinite number of dynamically interacting 
topological features (primitive particles). 

\section{The field of the primitive particle}
\label{field}

\begin{figure}[htb]
\begin{turn}{-90}
\centering 
\epsfysize=2cm
\includegraphics[scale=0.8]{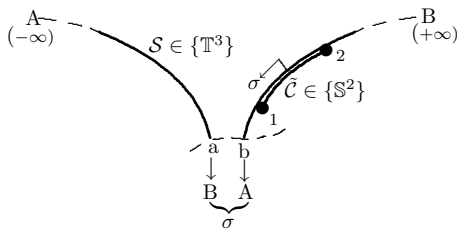}
\end{turn}
\caption{The ``throat'' (topological feature) $\sigma$ of 
a hypertorus $\mathbb{T}^3$ or the Klein bottle $\mathbb{K}^3$ and 
a control sphere $\tilde{\mathcal{C}} \in \mathbb{S}^2$ 
placed at $\sigma$ by sliding $\tilde{\mathcal{C}}$ within 
$\mathcal{S}$ towards the topological feature.
}
\label{fig:twochoices}
\end{figure}

\begin{figure}[htb]
\begin{turn}{-90}\epsfig{figure=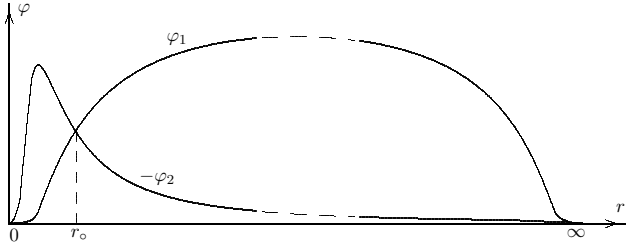, 
 width=3.5cm}\end{turn}
\caption{Two components, $\varphi_1$ and $\varphi_2$, of the equilibrium
field corresponding to the innermost and outermost points of 
the control sphere ${\Tilde{\mathcal{C}}}$.
}
 \label{fig:modifiedField}
\end{figure}

By regarding the described topological features of the manifold as primitive 
particles we have to assume the possibility of these particles 
interacting with each other; i.e., exercising a force on each other,
which in our case is realised by fluid pressure and tension. 
The corresponding field, $\varphi$, can be defined in terms of 
the flow density through an arbitrary 2-surface $\mathcal{C}$ surrounding 
$\sigma$. Since the inflow\,/\,outflow of test particles is isotropic, 
this field is spherically symmetric and roughly inversely proportional 
to the squared distance between $\mathcal{C}$ and $\sigma$ (for a discussion 
of the inverse-square law and scalar fields corresponding to solitonic 
objects see, e.g., \cite{birkhoff43} or \cite{ellis73}).
However, it is topologically impossible to enclose the central opening ($\sigma$) 
of a 3-torus with an arbitrary 2-surface.  
Due to this problem, the definition of $\varphi$ in our case
has to be modified. Let $\Tilde{\mathcal{C}}$ be a 2-sphere of radius 
$R_{\Tilde{\mathcal{C}}} \in (0,\infty)$ on $\mathcal{S} \in \{\mathbb{T}^3\}$ 
or $\{\mathbb{K}^3\}$ with its centre at 
$\sigma$ (but not enveloping $\sigma$), as shown in 
Fig.\ref{fig:twochoices}; namely, by sliding $\Tilde{\mathcal{C}}$ along $\mathcal{S}$ 
towards $\sigma$. 
The total flow $\Tilde{Q}$ through $\Tilde{\mathcal{C}}$ is nil since 
$\sigma$ is not enveloped by $\Tilde{\mathcal{C}}$. 
However, $\sigma$ divides $\Tilde{\mathcal{C}}$ into two hemispheres, so that 
the flow $\Tilde{Q}$ is split into two parts corresponding to 
the outermost and innermost hemispheres of $\Tilde{\mathcal{C}}$. 
By calculating the flow density at the innermost ``1'' and  outermost 
``2'' points of $\Tilde{\mathcal{C}}$ (Fig.\ref{fig:twochoices}) one can resolve 
the field $\varphi$ into two components:
\begin{equation}
\varphi(r)=\varphi_1(r) + \varphi_2(r),
\label{eq:superpos}
\end{equation}
where $r$ is the radial distance in spherical coordinates with the origin at $\sigma$.
Although the flow-line density grows as 
$R_{\Tilde{\mathcal{C}}}$ decreases, this density will
have a minimum at the origin (or will even vanish in the case of 
a non-compact manifold) since the boundary of the 
origin $\sigma$ is identified here with the outer (larger) circumference
of the torus.
The corresponding boundary condition (for a non-compact manifold) will be  
\begin{equation}
\varphi_1(0)=\varphi_2(0)=0.
\label{eq:boundary1}
\end{equation}
Then it follows that at some distance from 
$\sigma$ the density of flow-lines must have a maximum,
which would supposedly correspond to the upper bound of the manifold's 
curvature.
When changing $R_{\Tilde{\mathcal{C}}}$ (say, from zero 
to a maximal value), the points ``1'' and ``2'' of ${\Tilde{\mathcal{C}}}$ 
take opposite paths on the manifold:
\begin{equation}
\begin{matrix}
\text{``1''}: ~~~A \rightarrow a \\
\text{``2''}: ~~~a \rightarrow A 
\end{matrix}
\label{eq:paths}
\end{equation}
which would result in an asymmetry between the two components of the field; 
namely, $\varphi_2(r)$ will grow from a minimum (zero) to some maximal value, then 
decaying again to a minimum at the end of the path of the point ``2''
(Fig.\,\ref{fig:modifiedField}). The component $\varphi_1(r)$ will uniformly 
grow for most of the path of ``1''. Of course, eventually, 
at the end of this path $\varphi_1$ will decay to a minimum (zero).
Then, the second boundary condition for a non-compact manifold will be 
\begin{equation}
\varphi_1(\infty)=\varphi_2(\infty)=0.
\label{eq:boundary2}
\end{equation}

One should also take into account torsion
which, in the case of a 3-manifold, has three degrees of freedom 
\cite{ritis83},
and which would lead to the nonlinear Ivanenko--Heisenberg equation 
\cite{ivanenko38} and nonabelian degrees of freedom. The corresponding field 
would have a topological quantum number -- the colour analogy of 
helicity in fluid dynamics \cite{jackiw00}.
In the three-dimensional case there are six possible helical flow 
orientations, as distinct from, for instance, 
two-dimensional flow, with its bipolar vorticity and only four 
possible flow orientations 
\cite{tornkvist97}. 
Then, one can devise a force between primitive particles 
by regarding, for example, the vortex flow lines as electric 
currents 
\cite{baburova97} 
and calculating the Lorentz force 
between these currents. 
It follows that in the case of two like-charged 
(say, with inflows) primitive particles
of opposite vorticities the force $\varphi_1$ between the flow lines 
is attractive and $\varphi_2$ is repulsive. If the particles have like-vorticities 
the force $\varphi_1$ will be repulsive.

This accords with the
known pattern of attraction and repulsion between colour charges
\cite{suisso02}: two like-charged but unlike-coloured particles are attracted,
otherwise they repel. We can see that the colour pattern is readily
understood in terms of flow vortices on a manifold with
the $\mathbb{T}^3$ or $\mathbb{K}^3$ topology. 
The approximate antisymmetry of $\varphi_1$ and $\varphi_2$ in the 
vicinity of the origin implies that there should exist an equilibrium distance, 
$r_\circ>0$, such that 
\begin{equation}
\varphi_1(r_\circ)=-\varphi_2(r_\circ),
\label{eq:equilibrium2}
\end{equation}
in which case the fields cancel each other.
This breaks the initial spherical symmetry of the field.
For example, in the simplest case of a two-body system, the ground state
is the dipole whose orientation in space determines a preferred 
direction. 
The symmetry of scale invariance will also be broken because the equilibrium
distance, $r_\circ$, fixes the preferred scale unit for any preon-based 
composite system. 

\section{The simplest bound states of colour preons}
\label{hierarchy}

Let us consider some simple structures based on the fields (\ref{eq:superpos}) with
the boundary conditions (\ref{eq:boundary1}) and (\ref{eq:boundary2}) and the equilibrium 
condition (\ref{eq:equilibrium2}).
The following functional form
for the components of this field:
\begin{equation}
\begin{matrix}
\varphi_1(r)=&\varkappa\exp(-\kappa r^{-1}) \\
\varphi_2(r)=&{-\varphi'_1(r)~~~~~~~~~~~~~~~}
\end{matrix}
\label{eq:exp1}
\end{equation}
satisfies the conditions (\ref{eq:boundary1}) and (\ref{eq:equilibrium2}).
It does not match the second boundary condition (\ref{eq:boundary2})
but this does not matter for the cases with typical distances
between particles of the order of a few units of $r_\circ$.
The basic field does not necessarily have to be of the simplest
form (\ref{eq:exp1}). Its main feature (the capability of generating 
equilibrium particle configurations) could be derived from various 
physical considerations, including the geometry and shape
of the manifold, as was shown in \cite{ansoldi07}. 
It is also possible to satisfy the boundary condition (\ref{eq:boundary2}) 
by modifying the first component of the field in such a way as 
to nullify it at infinity, e.g.,
\begin{equation}
\tilde\varphi_1(r)=r^{-1}\varphi_1(r).
\label{eq:modifiedexp1}
\end{equation}
Nevertheless, for short distances we can adopt the field (\ref{eq:exp1}) 
as a simple example to illustrate the functionality of our model. 
This field corresponds to the following potential:
\begin{equation}
V(r)=(1-r)\exp(-\kappa r^{-1})-{\rm Ei}(-\kappa r^{-1}).
\label{eq:spotential}
\end{equation}
For simplicity (and to avoid any free parameters) the range 
coefficient $\kappa$ can be set to
unity. The coefficient $\varkappa=\pm1$ in (\ref{eq:exp1}) denotes the 
polarity of the field and must be chosen such as to reproduce the above
pattern of long-range attraction and short-range repulsion between particles.

A field of this kind reveals a potential surface with multiple local minima 
leading to kinematic constraints of a topological nature. This is 
analogous to the cluster formation scheme in molecular dynamics \cite{osenda02}
with the only difference that here we have to deal with the tripolarity of the 
fields. These constraints determine a unique set of
clusters, the simplest of which are dipoles
and tripoles composed of, respectively, two and three colour-preons. 
Obviously, the dipoles are deficient in one colour, whereas the tripole fields 
are colourless at infinity and colour-polarised nearby, which allows axial 
(pole-to-pole) coupling of these structures into strings. 
\begin{figure}[htb]
\begin{turn}{-90}\epsfig{figure=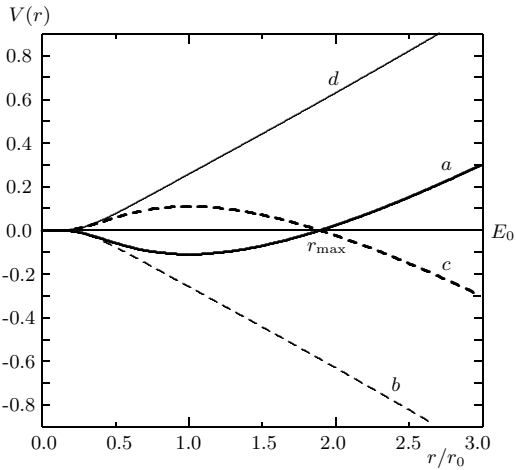, 
 width=7.0cm}\end{turn}
\caption{Equilibrium potential $V(r)$ 
corresponding to a colour-dipole system;\,\, ($a$)\,two
like-charged particles with unlike colours; 
($b$) the same but with like-coloured particles; 
($c$)\, oppositely charged particles with 
like colours; and ($d$) oppositely charged particles with 
unlike colours.
 \label{fig:spotential}
}
\end{figure}

The potential (\ref{eq:spotential}) is graphically shown in
Fig.\,\ref{fig:spotential}\,$a$.  
This is a typical double-well potential,
which is known to lead to chaotic oscillations and stochastic resonances 
\cite{wiggins89}. In cosmology, it results in the 
formation of domain walls during the phase transitions in the 
early universe. 
This potential is also known to be self-calibrated; i.e., it establishes 
the length, time and energy scales by, respectively, 
the separation between wells, oscillation frequency in the inverted 
barrier and barrier height. 
We shall use the half-separation between wells
(the equilibrium distance, $r_\circ$, 
corresponding to the minimum of the potential) as the 
basic unit length for this model. 
The speed unit, $v_\circ$, is also intrinsic to the potential 
(\ref{eq:spotential}) in the sense that, given the initial energy,
$E_0$, one can calculate the corresponding speed of a particle
moving in this potential.  
For example, the speed of a two-preon system 
with the potential (\ref{eq:spotential}) is calculated as 
\begin{equation}
v(r)=\sqrt{2(E_0-V(r))/\hat{m}},
\label{eq:speed2particles}
\end{equation}
where $\hat{m}^{-1}=m_1^{-1}+m_2^{-1}$
is the reduced mass ($\hat{m}=\frac{1}{2}$ since we use $m_1=m_2=1$). 
By setting the energy $E_0$ to some level, say, to zero 
at $r=r_{\rm max}$ we can calculate the maximal speed, 
$v_{\rm max}=v(r_\circ)\approx 0.937v_\circ$, 
expressed in units defined by the amplitude and range
coefficients of the function (\ref{eq:exp1}).
This also establishes the time scale by defining a unit time 
interval, $t_\circ$, such that $v_\circ t_\circ = r_\circ$, 
as well as the other necessary 
units, like, e.g., the unit of angular momentum, $L_\circ$,  
which would correspond to a preon of unit mass, $m_\circ$, moving 
with unit speed, $v_\circ$, along a circular path with unit 
radius, $r_\circ$.
The mass unit, $m_\circ$, can be defined by
calculating the energy of the field $\varphi$ for a single preon:
\begin{equation}
m^2= \frac{1}{2\pi}\,\int\limits_{\mathcal{S}} \varphi^2 \,d\mathcal{S} \,=\,1 
\label{eq:masssquared}
\end{equation}
with the first component of the field, $\varphi_1$, 
removed because the integral 
$\int \varphi_1^2 \,d\mathcal{S}$ diverges
and, obviously, cannot be used for this purpose.
If we ignore for a moment the third colour and consider a two-body 
system 
\begin{equation}
d^+_{\rm y}=\sigma^+_{\rm r} \wr\, \sigma^+_{\rm g}
\label{eq:bound1}
\end{equation}
-- a charged colour dipole with energy $E_0=0$ -- we can see that 
the charges in this system are confined (oscillating) within the region 
$(0,r_{\rm max})$, where $r_{\rm max}\simeq 1.894r_\circ$ 
(see Fig.\,\ref{fig:spotential}).  
The symbol\, $\wr$\, between $\sigma^+_{\rm r}$ and $\sigma^+_{\rm g}$
in (\ref{eq:bound1}) simply indicates that these two components oscillate with 
respect to each other; and the index {\scriptsize ``y''} 
(yellow=$\overline{\rm blue}$) indicates that this structure 
is deficient in the blue-polarity
field, $\varphi_1^{\rm b}$, which means that the charged colour 
dipole $d^+_{\rm y}$ cannot actually exist in free states by having infinite energy 
(unless we neglect the third colour).
The net charge of $d^+_{\rm y}$ corresponds to $+2q_\circ$. 
Its mass, $m_{d^+_{\rm y}}$,  in the first approximation, is equal to $2m_\circ$,
which is the exact value for the case when the centres of 
$\sigma^+_{\rm r}$ and $\sigma^+_{\rm g}$ coincide
and the components $\varphi_1^{\rm r}$ and $\varphi_1^{\rm g}$ 
of the field cancel each other exactly (the component $\varphi_1^{\rm b}$ being ignored).   

\begin{figure}[htb]
\begin{turn}{-90}\epsfig{figure=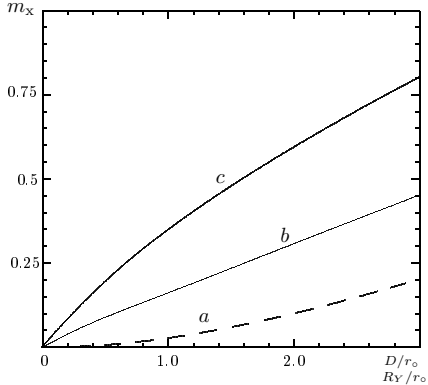,width=6.0cm}\end{turn}
\caption{Mass excess (in units of $m_\circ$) for some simple preon systems 
(normalised to the number of preons in the system) as a function 
of distance $D$ between preons. 
(a):\, A system of two like-charged preons with unlike 
colours; (b):\, two unlike-charged 
preons with unlike colours, and (c):\, three like-charged preons with 
unlike colours (in this case the abscissa corresponds to the tripole
radius, $R_Y$). 
}
\label{fig:massexcess}
\end{figure}

If the constituents of $d^+_{\rm y}$ are separated by some distance, $D$,
the fields $\varphi_1^{\rm r}$ and $\varphi_1^{\rm g}$ will cancel only
partially and the mass of this system will exceed the value $2m_\circ$ by a 
quantity (called hereafter the mass excess) which can be calculated 
by using (\ref{eq:masssquared}). The mass excess of the two-component 
system $d^+_{\rm y}$ as a function of distance between the components is shown 
in Fig.\,\ref{fig:massexcess}\,$a$. In its ground state
($D=r_\circ$) this system has a mass of $\approx 2.04\,m_\circ$.

For the system of two preons with like colours, say $\sigma^+_{\rm r} \, \sigma^+_{\rm r}$, 
we have to reverse the sign of $\varkappa$ in (\ref{eq:exp1}), 
which would result in a repulsive combined potential (curve $b$ in  Fig.\,\ref{fig:spotential})
having a maximum at the origin. This means that the system 
$\sigma^+_{\rm r}\sigma^+_{\rm r}$ 
cannot be formed in principle, even by neglecting the third colour.

The wells of the potentials corresponding 
to the preon pairs with opposite electric 
charges (curves $c$ and $d$ in Fig.\,\ref{fig:spotential}) imply that   
the neutral colour dipoles 
\begin{equation}
d^0_{\rm y}=\sigma^+_{\rm r} \wr\, \sigma^-_{\rm g} \hspace{0.5cm}
{\rm and} \hspace{0.5cm}
d^0_{\rm r}=\sigma^+_{\rm r} \wr\, \sigma^-_{\rm r}
\label{eq:bound3}
\end{equation}
could, in principle, be formed if the third colour were neglected. 
Then, the system $d^0_{\rm y}$ would have a vanishing mass
(100\% mass defect) for $D=0$ because in this case not only do the 
fields $\varphi_1$ of its two constituents have opposite signs 
and cancel each other, but the fields $\varphi_2$ also do. For $D>0$ 
the mass of $d^0_{\rm y}$ will be growing almost linearly with 
distance, as shown in Fig.\,\ref{fig:massexcess}\,$b$.

\begin{figure}[htb]
\begin{turn}{-90}\epsfig{figure=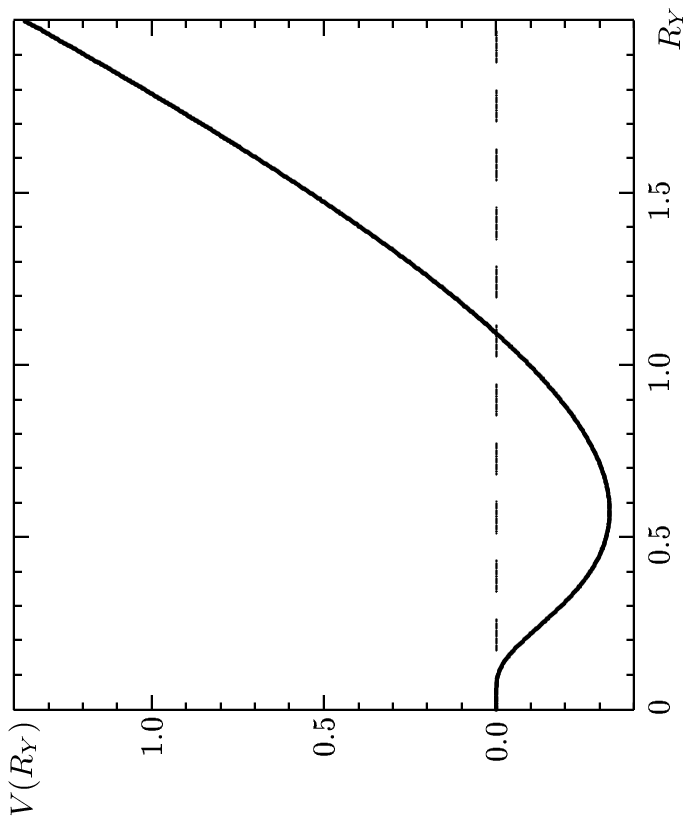,width=6.0cm}\end{turn}
\put(80,-10){
 \makebox(0,0)[t]{\begin{turn}{-90}
\epsfig{figure=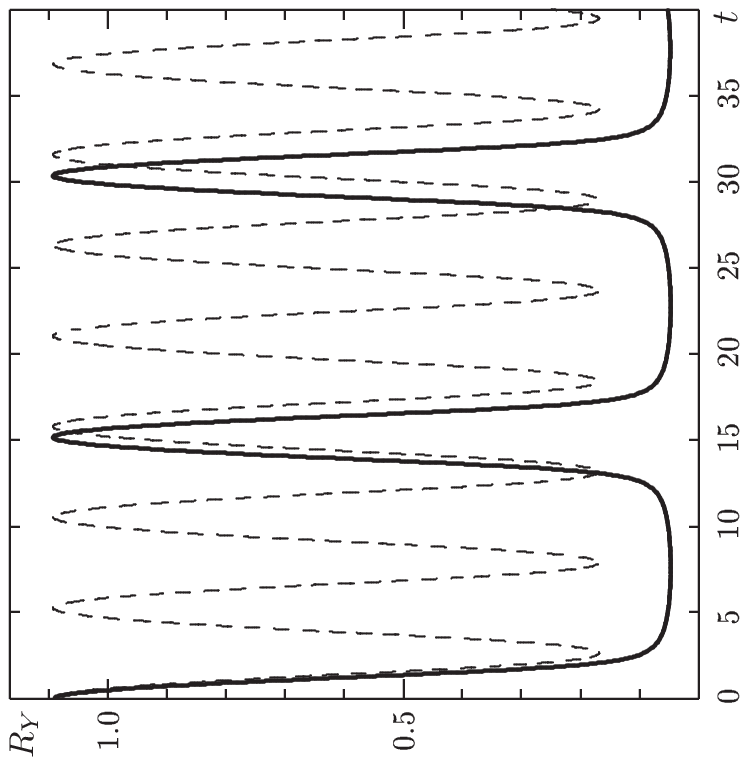,width=5.5cm}
\end{turn}
}}
\put(-95,-17){\makebox(0,0)[t]{a}}
\put(80,-17){\makebox(0,0)[t]{b}}
\put(-10,-120){\makebox(0,0)[t]{\scriptsize{$E_\circ$}}}
\put(-75,-126){\makebox(0,0)[t]{\scriptsize{$R_Y^{\rm max}$}}}
\caption{(a)\,Potential energy of the tripole system 
$Y$ as a function of its radius $R_Y$;\,\, (b)\,The pattern
of radial oscillations of a tripole 
with invariable mass (solid curve), and with 
the mass excess function taken into account (dashed curve). The amplitude 
of the oscillating $R_Y$ corresponds to the energy level 
$E_\circ=0^{\text{-}}$. Here time is in 
units of $t_\circ$ and the radii are in units of $r_\circ$.
}
\label{fig:potential2}
\end{figure}
The colourless bound state of three like-charged preons with 
complementary colours (the tripole)
\begin{equation}
Y^+=\sigma^+_{\rm r} \wr\, \sigma^+_{\rm g} \wr\, \sigma^+_{\rm b} \hspace{0.5cm}
{\rm or} \hspace{0.5cm}
Y^-=\sigma^-_{\rm r} \wr\, \sigma^-_{\rm g} \wr\, \sigma^-_{\rm b}\,,
\label{eq:bound4}
\end{equation}
has a finite mass.   
Given $E_0=0^{\text{-}}$, the tripole's radius, $R_Y$, will oscillate between zero 
and $R_Y^{\rm max} \simeq 1.09r_\circ$,
see Fig.\,\ref{fig:potential2}. 
The tripole system must be stable because it would take an infinite amount 
of energy to remove any of its three constituents from the system. In their ground state, 
these constituents will be separated from each other by the distance $D=r_\circ$, 
corresponding to the tripole's equilibrium radius,  
$R_Y=r_\circ/\sqrt{3}$, which minimises
the potential shown in Fig.\,\ref{fig:potential2}.  
Since the centres of the tripole constituents 
do not coincide, the fields $\varphi_1$ are cancelled only partially and 
the mass of the tripole's ground state has an excess of about $0.199\,m_\circ$ per 
preon over the mass of the state with $R_Y=0$ (curve $c$ in  
Fig.\,\ref{fig:massexcess}).

\section{Two- and three-component strings of tripoles}

\begin{figure}[htb]
\hspace{-1.5cm}
\begin{turn}{-90}\epsfig{figure=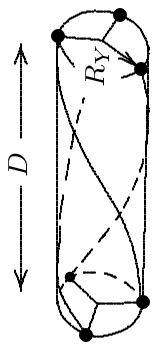, 
 width=3.5cm}\end{turn}
\put(50,10){
 \makebox(0,0)[t]{\epsfig{figure=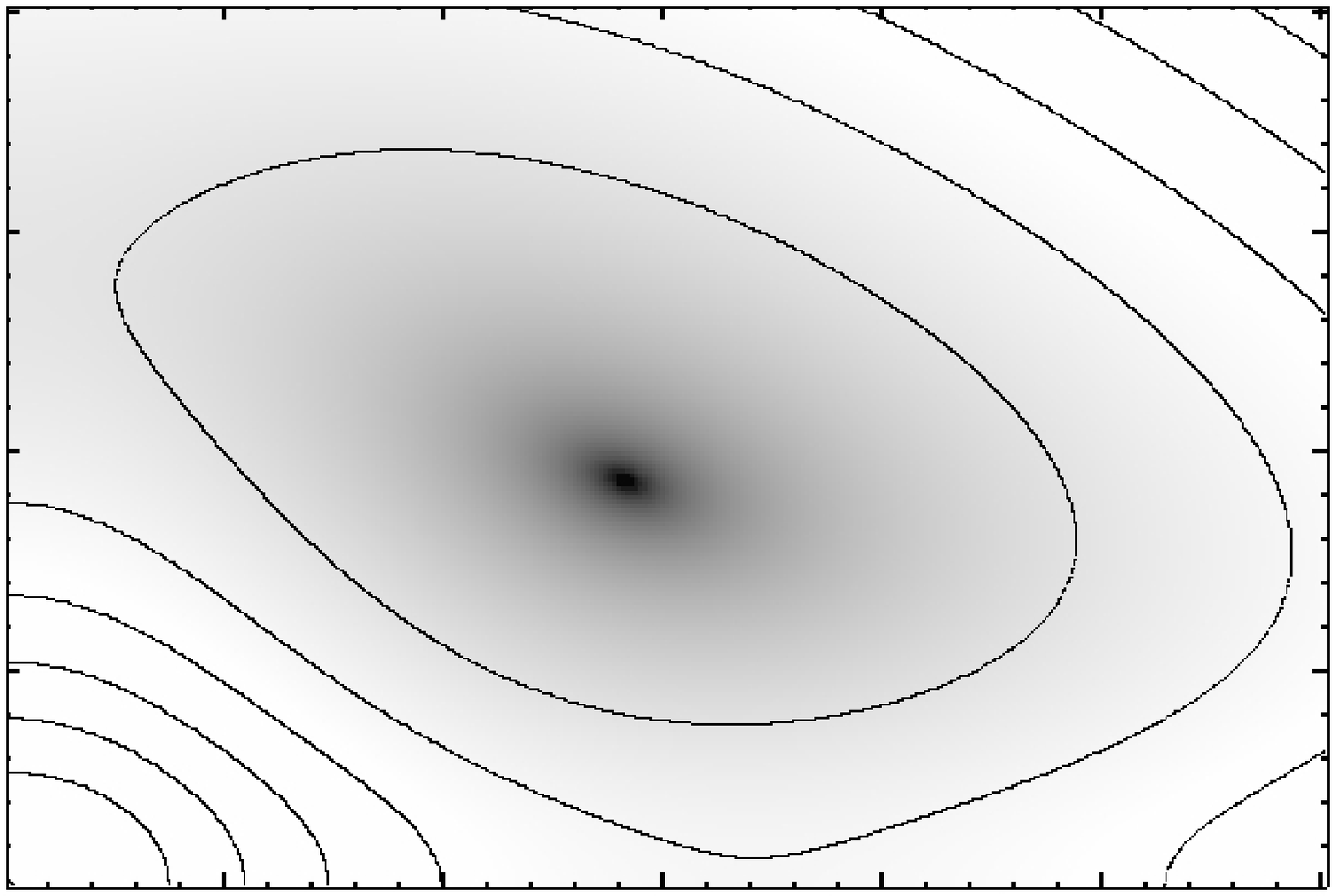,width=5.5cm}
 \put(-159,100){
 \makebox(0,0)[t]{\scriptsize{$R_Y$}}}
 \put(-10,5){
 \makebox(0,0)[t]{\scriptsize{$D$}}}
 \put(-152,5){
 \makebox(0,0)[t]{\footnotesize{$0$}}}
 \put(-105,5){
 \makebox(0,0)[t]{\footnotesize{$1.0$}}}
 \put(-58,5){
 \makebox(0,0)[t]{\footnotesize{$2.0$}}}
 \put(-160,34){
 \makebox(0,0)[t]{\footnotesize{$0.5$}}}
 \put(-160,59){
 \makebox(0,0)[t]{\footnotesize{$1.0$}}}
 \put(-160,82){
 \makebox(0,0)[t]{\footnotesize{$1.5$}}}
}}
\put(220,10){
 \makebox(0,0)[t]{\epsfig{figure=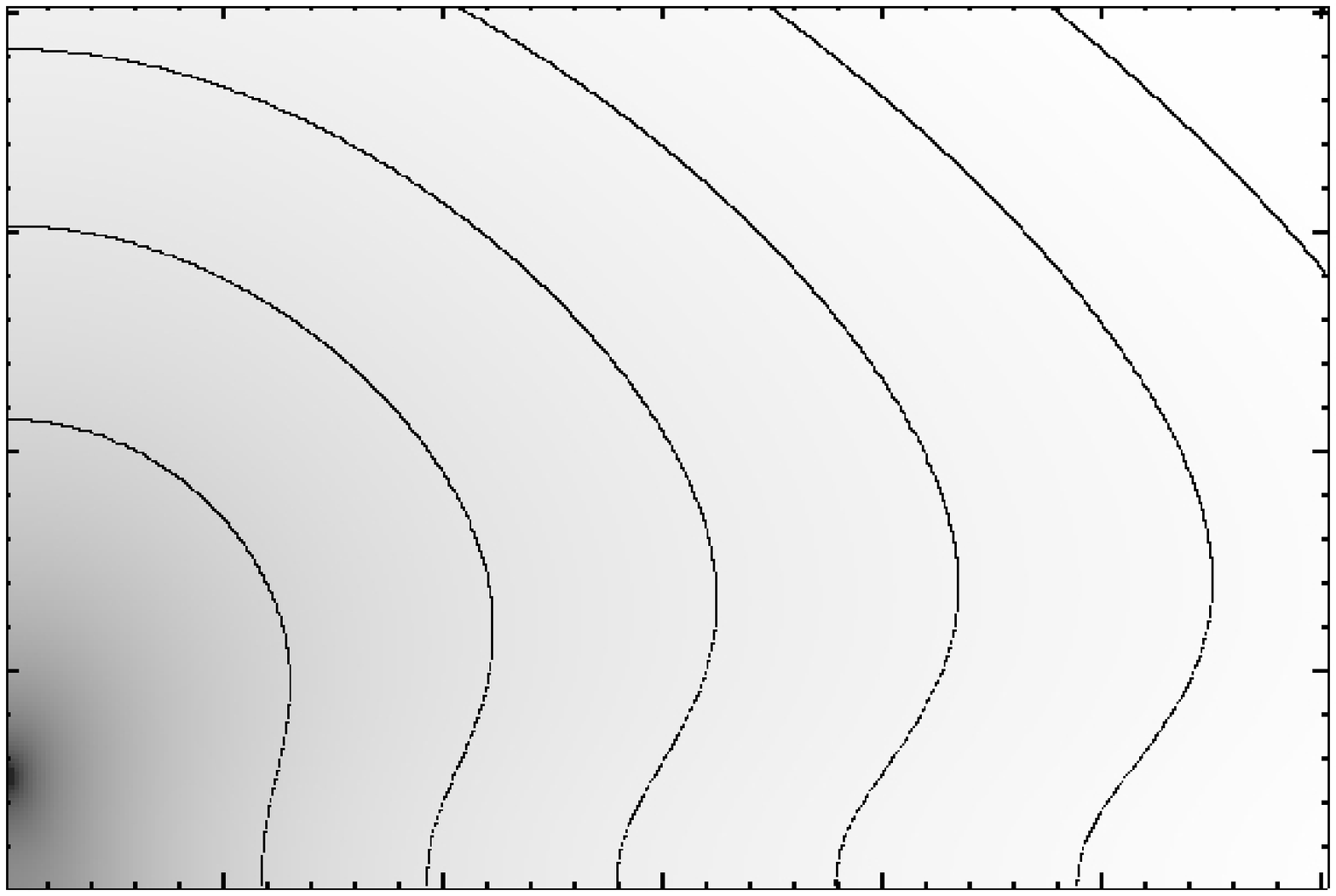,width=5.5cm}
 \put(-161,100){
 \makebox(0,0)[t]{\scriptsize{$R_Y$}}}
 \put(-10,5){
 \makebox(0,0)[t]{\scriptsize{$D$}}}
 \put(-152,5){
 \makebox(0,0)[t]{\footnotesize{$0$}}}
 \put(-105,5){
 \makebox(0,0)[t]{\footnotesize{$1.0$}}}
 \put(-58,5){
 \makebox(0,0)[t]{\footnotesize{$2.0$}}}
 \put(-160,34){
 \makebox(0,0)[t]{\footnotesize{$0.5$}}}
 \put(-160,59){
 \makebox(0,0)[t]{\footnotesize{$1.0$}}}
 \put(-160,82){
 \makebox(0,0)[t]{\footnotesize{$1.5$}}}
}}
 \put(45,9){
 \makebox(0,0)[t]{a}}
 \put(220,10){
 \makebox(0,0)[t]{b}}
\caption{Potential energy of a string formed of two tripoles $Y$ separated by 
distance $D$ and having radii $R_Y$ (scheme on the left); (a)\, two like-charged tripoles; 
(b)\, two oppositely charged tripoles.  
}
\label{fig:doubleTripole}
\end{figure}

Due to colour-polarisation of the tripole's fields, different 
tripoles can interact with each other
and form bound states. 
The simplest of them is a two-component string formed 
of either two like-charged tripoles: 
\begin{equation}
\gamma^+=Y^+ \wr\, Y^+
\,\,\,{\rm or} \hspace{0.5cm}
\gamma^-=Y^- \wr\, Y^-\,,
\label{eq:bound5}
\end{equation}
or oppositely charged tripoles:
\begin{equation}
\gamma^\circ=Y^+ \wr\, Y^-\,.
\label{eq:bound6}
\end{equation}
Obviously, the tripoles in these systems will 
be joined pole-to-pole to each other, rotated by $180^\circ$ (see the scheme on the left
of Fig.\,\ref{fig:doubleTripole}). 
The corresponding potential energies for the pairs of like- and unlike-charged
tripoles as functions of tripole's radii, $R_Y$, and separation, $D$, between 
the tripoles are shown in Fig.\,\ref{fig:doubleTripole}\,(a) and (b), respectively.  

The minimum of the potential for the second (electrically neutral) 
system $\gamma^\circ$, Fig.\,\ref{fig:doubleTripole}\,(b), 
is very close to the origin, which means that both
electric and colour fields in this system are cancelled almost entirely. 
Therefore, in its ground state this system will have a vanishing mass 
and will not be able to interact with other similar (neutral) systems,
so that in each point of space one can place as many 
of them as one pleases (the population density 
of these particles is not derivable from first principles). By being massless, 
these particles will move with maximal 
available speed in all possible directions, 
which can be regarded as an ideal gas of neutral particles. 
However, in the vicinity of an electric or colour
charge the constituents of $\gamma^\circ$ will be 
polarised either electrically or chromatically, 
converting this particle to an electric or colour dipole.
This property of $\gamma^\circ$ is 
important because within this framework one can use the gas 
of these particles to represent a polarisable medium (vacuum) 
where the motions and interactions of all other particles take place.
The dipole-dipole interactions between the polarised $\gamma^\circ$-particles
will result in the formation of instantaneous preonic configurations of different 
complexity, which can be regarded as a source of new particles, provided the 
supplied energy is sufficient for the complete separation of the polarised 
components of $\gamma^\circ$ (otherwise, these configurations could be 
used for modelling virtual particles).   

It is worthwhile mentioning here that the particle $\gamma^\circ$ 
is the only possible form of matter that can exist during the initial phase
of the universe's expansion, when the available volume is still too small
to allow any larger structures to be formed (when the universe's radius 
is smaller than $\approx 0.5 r_\circ$). The motions of these particles 
will be stochastic because of the unstable stationary point of the 
potential at the origin. This picture is somewhat similar to the 
Raitio's model \cite{raitio80}, according to which the universe at a 
very early moment consisted of stable particles of minimal size and maximal 
mass, called ``maxons'', carrying only the electric and colour charges 
and whose bound states have vanishing masses. Raitio's model is in turn 
close to the idea of bound states of ``maximon'' particles proposed by Markov 
in 1965 \cite{markov65}. 

\begin{figure}[htb]
\begin{turn}{-90}\epsfig{figure=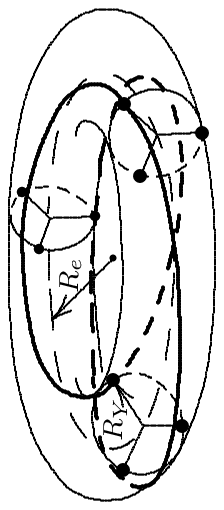, 
 width=3.5cm}\end{turn}
\put(100,10){
 \makebox(0,0)[t]{\epsfig{figure=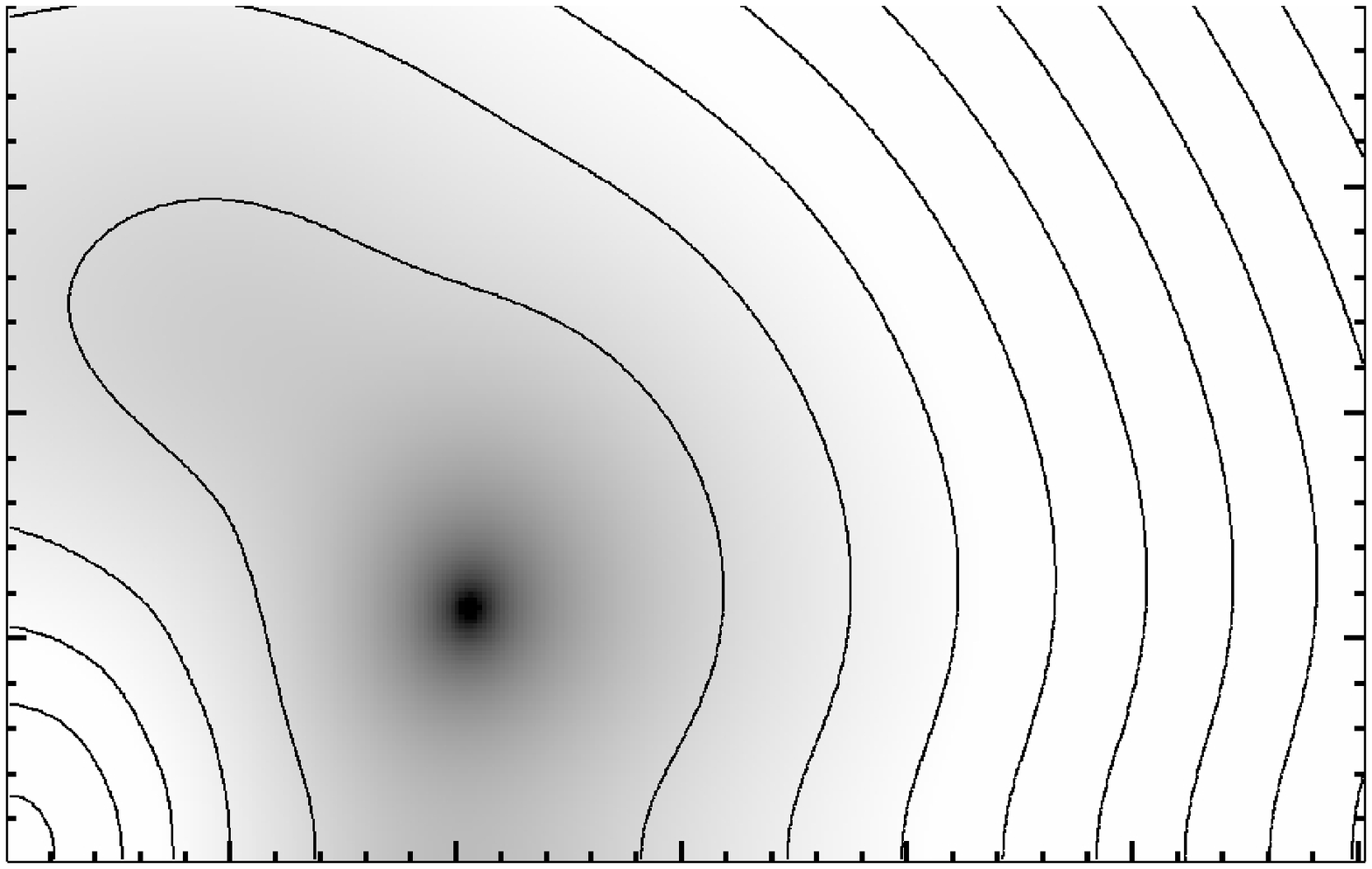,width=5.5cm}
 \put(-163,100){
 \makebox(0,0)[t]{\scriptsize{$R_Y$}}}
 \put(-3,4){
 \makebox(0,0)[t]{\scriptsize{$R_e$}}}
 \put(-155,2){
 \makebox(0,0)[t]{\footnotesize{$0$}}}
 \put(-105,2){
 \makebox(0,0)[t]{\footnotesize{$1.0$}}}
 \put(-56,2){
 \makebox(0,0)[t]{\footnotesize{$2.0$}}}
 \put(-164,34){
 \makebox(0,0)[t]{\footnotesize{$0.5$}}}
 \put(-164,59){
 \makebox(0,0)[t]{\footnotesize{$1.0$}}}
 \put(-164,82){
 \makebox(0,0)[t]{\footnotesize{$1.5$}}}
}}
\caption{Potential surface corresponding to 
the loop $e^\pm$ of radius $R_e$ formed of three like-charged tripoles $Y$ with radii $R_Y$ 
(scheme on the left). The radii are expressed in units of $r_\circ$. 
}
\label{fig:minimum}
\end{figure}

By analogy with the three-component system (\ref{eq:bound4}) we can consider 
a three-component string, $e^\pm$, formed of like-charged tripoles:
\begin{equation}
e^+=Y^+ \wr\, Y^+ \wr\, Y^+ \hspace{0.5cm}
{\rm or} \hspace{0.5cm}
e^-=Y^- \wr\, Y^- \wr\, Y^-\,.
\label{eq:bound7}
\end{equation}
 The potential energy of this string will be minimised for its loop-closed 
configuration with its constituent tripoles rotated by $120^\circ$
with respect to each other (see the scheme on the left of Fig.\,\ref{fig:minimum}). 
The minimum of the potential surface shown in Fig.\,\ref{fig:minimum} 
corresponds to the static equilibrium of this loop. 

\noindent
But, in fact, such a static state cannot be stable because the tripoles comprising the loop 
retain their rotational and translational degrees of freedom (around and along 
their common ring-closed axis) and will be moving under the influence of 
all the other existing particles, for example, under the stochastic action
of the above mentioned gas of the neutral two-component tripole 
strings.
The dynamical parameters of this system  
will be quite different from those corresponding to the static case 
shown in Fig.\,\ref{fig:minimum} because each of the spinning 
tripoles will generate a magnetic field modifying (and stabilising)  
the motions of the two other tripoles in the loop \cite{pati81}.  
By the standard dynamo mechanism these motions will induce 
toroidal and poloidal magnetic fields which in turn will 
maintain the rotational and orbital motions of the looped tripoles
(we shall discuss the dynamics of this system elsewhere).

The colour-currents corresponding to the motion of each individual preon 
in this system are helices (Smale-Williams curves) which, by their closure,
make a $\pi$-twist around the ring-closed axis of the structure
with either clockwise or anticlockwise winding 
(the scheme shown in Fig.\ref{fig:minimum} corresponds to the 
anticlockwise winding of the currents).
Such a twisting dislocation of the phase is a conserved quantity called
topological charge \cite{kleman83} or dislocation index, which has a sign 
corresponding to the winding direction (clockwise 
or anticlockwise) and the magnitude related 
to the winding number per $2\pi$-orbit path.
In these terms, the $\pi$-phase shift of the currents in the 
structure $e^\pm$ corresponds to a topological charge $S=\pm \frac{1}{2}$,
identifiable also with the particle internal angular momentum (spin).

It is seen that the path of each preon belonging to a particular tripole overlaps exactly
with the paths of two other preons that belong to two other tripoles 
of the structure and whose colour charges are complementary to the 
colour charge of the first preon. That is, the currents that form 
the structure $e^\pm$ are dynamically colourless
and, averaged in time, the field of this particle will have only
two polarities corresponding to the conventional electric field.

\section{A hierarchy of preonic bound states}

One finds that, besides the simplest bound states of colour preons 
discussed in the previous sections, the field (\ref{eq:exp1}) gives rise to 
a rich variety of other structures.
For example, a string of tripole-antitripole pairs has a symmetry similar to
that of the structure (\ref{eq:bound7}), which allows its 
closure in a (minimal) loop, $\nu_e$, containing six such pairs (twelve tripoles). 
Like $e^\pm$, this structure is also dynamically colourless but,
unlike $e^\pm$, it is electrically neutral. Despite its null electric and colour charges, 
the particle $\nu_e$ can, nevertheless, interact with  
$e^\pm$ through its residual (oscillating) chromoelectric fields, given that the motions 
of the constituents in both structures are synchronised. 
We shall discuss the issue of synchronisation elsewhere, leaving it here 
as a hypothetical possibility, although from simple considerations it is 
seen that this is likely to be the case. For example, by rotating one of the 
tripoles in the two-component string shown in Fig.\,\ref{fig:doubleTripole}, 
one can find that to maintain the equilibrium the second tripole will start rotating 
in the same direction and with the same angular speed as the first tripole. 
Likewise, if two initially static loops
$e^\pm$ (Fig.\,\ref{fig:minimum}), were placed with their 
orbital planes parallel to each other in a position of equilibrium (when each 
preon from one loop faces a pair of preons from the other loop, the latter two carrying the 
complementary colour-charges to the first one) then the rotation of one of the loops
will cause synchronous rotation of the other in order to minimise the combined potential
of the two structures.   
So, neighbouring structures $\nu_e$ and $e^\pm$ will,
indeed, interact with each other, the more so because, curiously enough, 
the helical patterns of their colour currents perfectly match each other, 
provided that the two particles have like-topological charges (spins), otherwise 
their residual chromoelectric fields will be repulsive. 
Two like-topological charges attract each other only 
if the particles carrying these charges were of two different kinds
with the distinct radii of their loops, allowing one of them passing 
through the central opening of the other (which is the case for the 
structures $\nu_e$ and $e^\pm$). Particles of the same 
kind and size (e.g., two particles $e^\pm$) will be repulsed from each 
other by having like-topological charges and attracted to each other
by having opposite topological charges.   
This feature of the tripole loops might shed some light on the origin 
of the Pauli exclusion principle.  

By their properties, which include spin and charge (see Section\,\,5), 
gyromagnetic ratio \cite{yershov07} and parity \cite{yershov05}, 
the three- and twelve-tripole loops ($e^\pm$ and $\nu_e$) can be identified with 
the electron and its neutrino, respectively.  
One can also find that different combinations of these loops (involving 
also the tripoles $Y^\pm$) constitute a hierarchy of structures 
identifiable with the observed variety of elementary particles
(we shall discuss this in detail elsewhere).
Due to the topological constraints, the number of constituents in each structure is
well-defined by the minimum of its potential energy, so that the hierarchy
turns out to be unique.

The masses of these structures ensue from the energies of 
the motions of their constituents less binding energies (mass defects), 
which are known as standard mass-generating mechanisms for the composite systems. 
Therefore, within the framework of this model, there is no need of extra fields, 
like Higgs, to generate particle masses. 
The same null prediction can be made with respect to the supersymmetric 
partners of the known particles, since this model does not generate more 
structures than are needed to explain the origin of the observed variety of particles. 

As was mentioned in Section\,\,4, the fields of opposite polarities (or complementary colours) 
almost entirely cancel each other if two unlike-charged particles (structures) 
overlap or get close to each other, which also nullifies the masses of 
the neutral loops, like $\nu_e$. This accounts for the so-called 
``mass paradox'', according to which the momentum uncertainty 
for the constituents of a structure confined in a small volume of 
a composite particle should necessarily be greater than the mass of 
the host particle they comprise. In our case the almost infinite energies 
of the primitive particles are nullified by their binding energies,
which is the case for many composite models \cite{silveira85}. 

The binding energy, $E_Y$, between tripoles in a composite structure 
is due to residual effects of the chromoelectric fields of the preons 
constituting these tripoles. Therefore, $E_Y$
must be smaller than the binding energy between the preons constituting
the tripole Y. 
To get an idea about the order of magnitude of $E_Y$ we can 
take Dehmelt's estimate of the electron radius, $R_e=10^{-22}$~m, 
obtained by extrapolating the experimentally measured $g$-factor-to-radius 
ratios for the known near-Dirac composite particles (see \cite{dehmelt89} 
for details). Of course, this value of $R_e$ is unrelated to the classical 
electron radius or the Compton radius because it was obtained by assuming 
that the anomalous magnetic moment of the electron is due to its 
compositeness. Thus, $R_e$ in this case corresponds to the physical size 
of a structure forming the composite electron.
If we assume that $R_e$ corresponds to the radius of the 
loop $e^\pm$ in Fig.\,\ref{fig:minimum} then the distance $D_Y$ between 
each pair of the tripoles constituting this loop will be roughly
twice the value of $R_e$. The binding energy between the tripoles
must exceed (or at least be equal to) ${5\cdot 10^3}$~TeV, which is
the energy scale corresponding to the distance $D_Y=2 \cdot 10^{-22}$~m, 
that is, $E_Y \ge 5\cdot 10^3$~TeV. 
This is far greater than the energy scales accessible
by modern experimental techniques. For example, the proton beam energy 
of CERN's Large Hadron Collider (LHC) is expected to be of about 7~TeV 
\cite{dissertori05}. 
%
This seems to result in very few 
(if any) new phenomena observable with the use of the most powerful modern 
colliders. 

Nevertheless, it is possible that some phenomena,
either supporting or falsifying our model, could be seen at the TeV energy 
scale. For instance, quark sub-structures might be revealed by the 
LHC in the form of deviations from the QCD expectation of the high energy part of the jet 
cross-sections \cite{stump03} and also in hadronic di-jets angular correlations 
\cite{treille06}. 

As was mentioned in Section~\ref{field}, our model is based solely on the 
symmetry SU(3)$\times$U(1) of the basic field, whereas the weak 
interaction is viewed  as a low-energy property of composite systems. 
That is, the weak interaction in our model
is not fundamental but a residual effect of underlying chromoelectric 
fields, similar to the Yukawa interaction between nucleons,
which is thought to be a residual interaction between quarks inside 
each individual nucleon. This is a situation typical for most composite
models \cite{senju85}. It is worth mentioning also that excluding the weak force 
from the fundamental forces is not at all unreasonable from a theoretical point of view and does not 
create any problem in the rest of physics, as was recently shown by 
Harnik, Kribs and Perez \cite{harnik06}. Then it follows that the weak gauge
bosons must also be composite, which might lead to unusual observations in the TeV 
energy range within the reach of the LHC.
 
The composite nature of particles, e.g. of the electron, might also reveal 
itself at low energies through synchronisation between the constituents of 
the composite electrons in ferromagnetic electron liquids \cite{bloch29} 
and low-dimensional electron systems, like Wigner crystals \cite{fogler06}. 
Another possibility is to detect some exotic electron bound states, like those 
observed by Jain and Singh \cite{jain07} 
in their experiments at CERN aimed at searching for low-mass 
neutral particles decaying into e$^+$e$^-$ pairs. They have reported the
detection of short-lived neutral particles having masses of 
$7\pm{1}$~MeV and $19\pm{1}$~MeV. The former might correspond to the minimal
spherically closed shell of eight electron-positron pairs \cite{yershov05} 
comprising a mass of about 8\,\,MeV, less, of course, the binding energy between the 
constituents of this shell.

\section{Cosmological singularity}
\label{cosmologicalSingularity}

The preonic structures discussed in the previous sections 
correspond to large manifolds, whereas
for a reduced volume the situation must be quite different. 
Let us assume that our manifold $\mathcal{S}$ is evolving 
(either expanding or contracting), with its curvature changing in time. 
During the contraction phase the growing curvature will reach 
its upper limit corresponding to the intrinsic curvature of the 
primitive particles. Since the potential (\ref{eq:spotential}) 
has a stationary point at the origin, all the primitive particles 
on $\mathcal{S}$ will be squeezed into a minimal volume, 
their centres coinciding in the origin.
At this stage the contraction rate and the curvature of 
the manifold are maximal, whereas its volume is minimal 
(but not vanishing).   
This structureless (matter-free) superposition of 
primitive particles corresponds to the appearance of a de Sitter
state, which is the appropriate candidate for a non-singular 
initial state of the universe \cite{gliner75,markov83} (for a review see
\cite{olive90}). 
 
Besides the problem of the cosmological singularity,
there are some other important cosmological problems that 
must be addressed by any model of the universe. 
Here we shall briefly outline the nature of these problems and the 
possible ways of addressing them within the framework of our model.

\vspace{0.2cm} 
{\it \underline{Entropy problem}. \,}
The entropy problem is related to the ``specialness''  of the initial 
state of the universe \cite{penrose82}.
The fact of the mere existence of the second law of thermodynamics implies 
that the initial state of the universe 
must have had a very low entropy value \cite{penrose89}. 
According to the second law of thermodynamics, in cyclic-universe models 
the universe must be increasing in size on each next cycle \cite{tolman31}. 
Extrapolated backwards in time, this leads to the same 
problem of the initial singularity that cyclic models try to resolve. To avoid this 
problem the extra amount of entropy has to be removed from the universe.
For example, in the oscillating model of Braun and Frampton
\cite{brown07} this is done by appealing to phantom dark 
energy and a deflation mechanism. The authors of this model propose that near 
the turnaround of an oscillating universe only one causal patch of the manifold 
is retained, the other patches contracting independently to separate universes. 
Thus, the entropy excess is permanently removed from  our universe when the 
scale factor is deflated to a tiny fraction of itself, and the universe starts
a new cycle of expansion containing vanishing entropy. 
In a recent cyclic-universe model by Biswas \cite{biswas08}, in each previous 
cycle the universe remains more and more time in the thermal equilibrium 
entropy-preserving phase (called ``Hagedorn soup'' \cite{hagedorn65}) and less  
time in the entropy-producing phase, which, if extrapolated to the infinite past, results in 
a constant entropy value. This makes Biswas' model non-singular 
and consistent with the second law of thermodynamics. 

Our model contains an even more radical solution to the entropy problem.
Namely, during the maximal contraction phase, when 
the primitive particles lose any degree of freedom for motion,  
the entropy of the collapsed universe is minimised (in fact, 
nullified), as opposed to the conventional view that entropy never 
decreases 
\cite{penrose79}. 
This simply reflects the fact that entropy, 
being a macroscopic statistical parameter, cannot be used for characterisation 
of a single-element microscopic system like the collapsed universe modelled 
with the fields (\ref{eq:exp1}).
Of course, the universe's entropy will be growing during both the expansion
and contraction phases, but at the end of the contraction phase the universe 
will invariably go to the above-mentioned highly-ordered single-element state, 
which means that after each cycle the universe will be completely renewed. This
circumvents the problem of the ever-increasing entropy 
characteristic of many cyclic-universe models. For the same reason
this model avoids the problem related to the enormous entropy production
in previous cycles due to black hole formation \cite{penrose90}. 
The possibility of a cold initial state of the universe was 
first discussed by Zeldovich and Novikov \cite{zeldovich75}. They have calculated
the present-day entropy that would correspond to this initial state and 
found that this entropy surprisingly well matches the observed value of 
about $10^9$ per baryon. In the case of the de Sitter initial state this
was first calculated in \cite{gliner75}.

\vspace{0.2cm} 
{\it \underline{Bouncing problem}. \,}
Lifschitz and Khalatnikov \cite{lifschitz63} discussed
the possibility for the worldlines of matter particles (in 
a non-homogeneous and non-isotropic collapsing universe) to cross in 
the same point but pass by one another and continue moving
without reversing their direction of motion, thus, making the universe 
{\sl to appear} to bounce back from the contraction. However, 
the prospects of this model were severely dimmed by Penrose and Hawking's
singularity theorems \cite{penrose69}. 
In our model the natural Lifschitz-Khalatnikov bounce can occur 
in the case of the $\mathbb{K}^3$-topology of the universe 
because this topology allows passing from the contraction to expansion 
phase with maximal speed and without reversing the
direction of motion of test particles (in this case the manifold simply turns ``inside-out''
without changing its topology). 
By contrast, a hypersphere $\mathbb{S}^3$ or the hypertorus $\mathbb{T}^3$ 
cannot pass homeomorphically  from the contraction to expansion phase, which 
leaves us with only one possibility: the topology of the universe 
must be $\mathbb{K}^3$.

\vspace{0.2cm} 
{\it \underline{Flatness problem}. \,}
Observations show that the universe seems to be spatially flat. 
This would be an extreme coincidence because a flat universe
is a special case.
In our model this problem is solved without a fine tuning 
or appealing to the anthropic principle
needed typically for treating this problem 
(see, e.g., \cite{tangherlini93,durrer96,biswas08} and 
for a recent review \cite{lahav04}). 
The manifold $\mathbb{K}^3$ is known to belong to the class of
Euclidean manifolds, which are finite and geodesically complete 
\cite{thurston84}, implying that any metric on a Klein bottle is conformally
equivalent to a flat metric \cite{jakobson06}. Moreover, in our case 
the global geometry of the spatial slice of the manifold should be very close 
to Euclidean at any stage of its evolution. This can be seen simply from the fact that in the
$\mathbb{K}^3$ case the manifold is turned ``inside-out'', with the curvatures
of the ``inner'' and ``outer'' parts of the manifold 
being equal by their magnitudes and opposite 
in sign. The total curvature would then be observed almost vanishing, with a residual 
effect due to the presence of matter (primitive particles and their combinations), 
so that our model implies a nearly flat universe, which is what is actually 
observed \cite{spergel03}. 

\vspace{0.2cm} 
{\it \underline{Charge asymmetry}. \,}
The model described is symmetric
with respect to the number of positive and negative charges 
(right and left parts of Fig.\ref{fig:blackhole}) as each such pair is viewed as   
a manifestation of the same topological feature. 
The charge asymmetry 
within the framework of this model could be understood in terms 
of the distinct paths that inflows and outflows of test particles
take on the ``inner'' and ``outer'' sections 
of the manifold. This is likely to lead to a difference in the energy content
of these two sections because, due to the existence of an upper limit on the 
manifold's curvature, these sections will never overlap exactly and, thus, will 
have slightly different volumes, except for the instantaneous configuration
corresponding to the maximal contraction phase.
The sign of the difference between their energies will be 
reversed after each recollapse of the manifold, recovering the CPT symmetry for 
the entire expansion-contraction cycle.  
  
\vspace{0.2cm}  
{\it \underline{Horizon problem}. \,}
There exist a series of models addressing the horizon problem 
(as well as some other cosmological problems) by assuming 
the variability of the fundamental speed \cite{albrecht99,chakraborty02}.
In our model the magnitude of the fundamental speed is presumably related
to the speed of the flow forming the spinning manifold
(see Section\,2)
In an expanding universe with rotation the value of rotation 
decreases \cite{goedel49}, and so too -- the fundamental speed,
which should have been maximal when the universe was passing 
from the contraction to expansion phase and minimal when the universe 
reaches its maximal size. 
In our case the variability of the fundamental speed 
can follow from the necessity to preserve the angular momentum 
of the spinning manifold \cite{fan50}.
So, the colour-preon model fully embraces the variable 
fundamental speed theory, which provides an alternative 
to the standard inflationary picture 
and leads to the predictions with respect 
to the homogeneity, flatness and perturbation spectrum 
similar to those made by the standard inflationary model
\cite{magueijo03}.  

\vspace{0.2cm} 
{\it \underline{Structure formation}. \,}
The potential (\ref{eq:spotential}), by having an unstable stationary point at 
the origin, implies spontaneous symmetry breaking, 
leading to the separation of colour-charges from each other
and to their self-organisation in complicated structures 
on a certain stage of the manifold expansion. 
The initial spherical symmetry of the field and the symmetry of scale
invariance will be broken when the size of the universe grows to a few 
units of $r_\circ$, i.e., when the primitive particles forming the
tripole structures pass from their unstable
initial state with energy $E_\circ$ and $R_Y=0$ (see Fig.\,\ref{fig:potential2})
to their ground state with the equilibrium radius $R_Y \approx 0.58\,r_\circ$.
The spherical symmetry is broken because each tripole acquires a
rotational axis, 
whereas the scale invariance is broken because the tripole constituents 
cannot be removed from this system (see Sect.\,\,4).

The further growth of the universe implies the formation of 
a complicated network of tripole strings and loops, 
as was shown in \cite{yershov05}. They act stochastically on each other and
on the spacetime metric, which is likely to affect the 
subsequent large-scale structure formation in the universe in the same 
manner as in the standard cosmology \cite{hagedorn65,omnes69}.

\section{Summary and Discussion.}

As we have seen, our model gives an insight both
into the nature of the cosmological singularity and the origin 
of elementary particles.
It has inbuilt units of length, time and speed, fixing a scale against which 
this model can be compared with observations.  Within its framework the 
dynamics of particles and sub-particles can be described in terms of 
well-established relativistic mechanics and Maxwell's electrodynamics.
Although the characteristic length scale of this model 
is likely to be close to the Planck-length, one can, in principle, 
use this model to make calculations for much larger structures 
(such as nucleons). So, an interesting area for further research 
would be exploring the phenomenological consequences of this framework 
on scales where more experimental tests could be 
conducted, e.g., on the scales of nuclear and atomic structures.  

\vspace{0.2cm}
One can find that, besides the problems mentioned in the previous section, 
the colour-preon model can shed light on the following 
issues, mostly related to particle and nuclear physics: 
\begin{itemize}
\item
the origin of the observed variety of particle species; 
\item
the neutrino left-handedness; 
\item
the electroweak symmetry breaking mechanism; 
\item
the origin of the Pauli exclusion principle;
\item
the indistinguishability of particles born in different reactions;
\item
the invariability of particle masses and charges; 
\item
the equivalence of the magnitudes of the proton and 
electron electric charges;
\end{itemize}
and some others.
The problem of the possible origin of the observed variety of particles
is related to the problem of the uniqueness of the universe \cite{susskind03}. 
As was mentioned in Section\,\,6, in our case the particular configuration 
and the number of constituents in each preonic cluster is 
well-defined by the minimum of the combined potential of this 
cluster. Therefore, the observed variety of particle species, as well as 
their internal structures, are determined uniquely, and the universe 
modelled with the use of the field (\ref{eq:superpos}) will generate 
always the same set of particles having the same properties. It is important 
to note that this set is pre-determined by the SU(3)$\times$U(1)-symmetry of 
this field, so that any particular choice of its functional form 
(like, for example, the exponential form used in this paper) will only change
the scale units, not altering the variety of particle species,   
We shall discuss the other of the above issues elsewhere (see also Sect.\,\,6 and Refs. 
\cite{yershov07}, \cite{yershov05}) noting here that it seems very 
attractive to address most of the fundamental questions of the 
standard particle physics and cosmology based on a few primary 
constituents.

\end{document}